\documentclass[a4paper,11pt]{article}
\pdfoutput=1 

\usepackage{jinstpub} 
\usepackage{graphicx}
\usepackage{hyperref}
\usepackage{subfig}

\newcommand{\gx}{\textsc{GlueX}}

\title{\boldmath Installation and Commissioning of the \gx{} DIRC}

\author[b,c]{A. Ali}
\author[e]{F. Barbosa}
\author[f]{J. Bessuille}
\author[e]{E. Chudakov}
\author[b]{R. Dzhygadlo}
\author[e,f]{C. Fanelli}
\author[d]{J. Frye}
\author[f]{J. Hardin}
\author[g]{A. Hurley}
\author[f]{E. Ihloff}
\author[a]{G. Kalicy}
\author[f]{J. Kelsey}
\author[g]{W. B. Li\thanks{wenliang.billlee@gmail.com}}
\author[f]{M. Patsyuk}
\author[b]{J. Schwiening}
\author[d]{M. Shepherd}
\author[g]{J. R. Stevens}
\author[e]{T. Whitlatch}
\author[f]{M. Williams}
\author[f]{Y. Yang}

\affiliation[a]{Catholic University of America\\Washington DC, United States}
\affiliation[b]{GSI Helmholtzzentrum f\"ur Schwerionenforschung GmbH\\Darmstadt, Germany}
\affiliation[c]{Goethe University Frankfurt\\Frankfurt am Main, Germany}
\affiliation[d]{Indiana University\\Bloomington, IN, United States}
\affiliation[e]{Jefferson Lab\\Newport News, VA, United States}
\affiliation[f]{Massachusetts Institute of Technology\\Cambridge, MA, United States}
\affiliation[g]{William \& Mary\\Williamsburg, VA, United States}
\emailAdd{wenliang.billlee@gmail.com}

\abstract{The \gx{} experiment takes place in experimental Hall D at Jefferson Lab (JLab). With a linearly polarized photon beam of up to 12 GeV energy, \gx{} is a dedicated experiment to search for hybrid mesons via photoproduction reactions. The low-intensity (Phase I) of \gx{} was recently completed; the high-intensity (Phase II) started in 2020 including an upgraded particle identification system, known as the DIRC (Detection of Internally Reflected Cherenkov light), utilizing components from the decommissioned BaBar experiment. The identification and separation of the kaon final states will significantly enhance the \gx{} physics program, by adding the capability of accessing the strange quark flavor content of conventional (and potentially hybrid) mesons. In these proceedings, we report that the installation and commissioning of the DIRC detector has been successfully completed.}

\begin{document}



\proceeding{International Conference on Instrumentation for Colliding Beam Physics\\
24 - 28 February 2020,\\ 
Budker Institute of Nuclear Physics and Novosibirsk State University (NSU), Novosibirsk, Russia}

\maketitle

\section{Introduction}
\label{sec:intro}

\begin{figure} [ph!]
	\centering
     \includegraphics[width=1\textwidth]{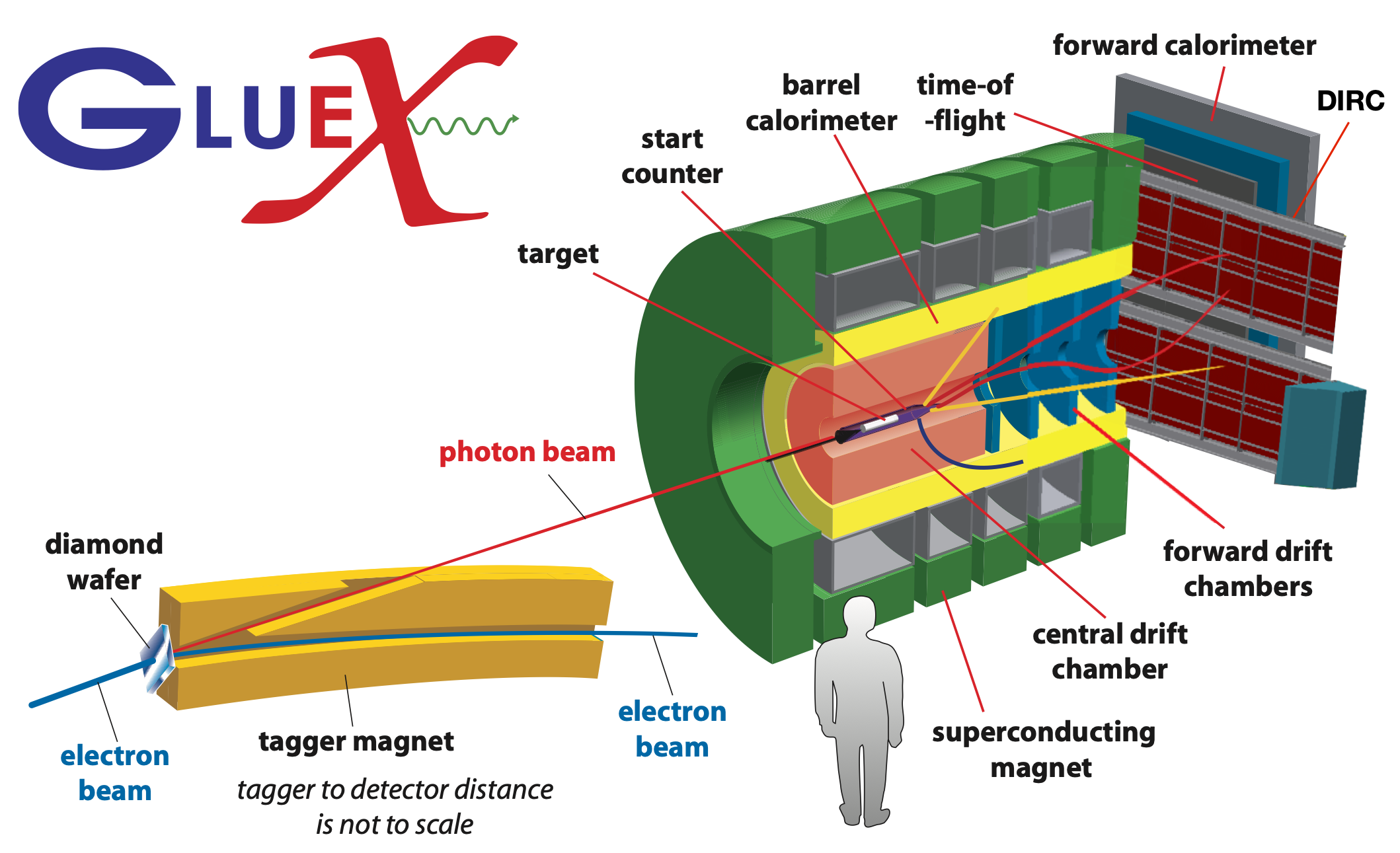}
     \caption{A schematic diagram of the Hall D beamline, tagger and \gx{} detector stack.  The DIRC detector is installed between the solenoid and the time-of-flight detector.}
\label{fig:GlueXcartoon}
	\centering
	\includegraphics[width=0.50\textwidth]{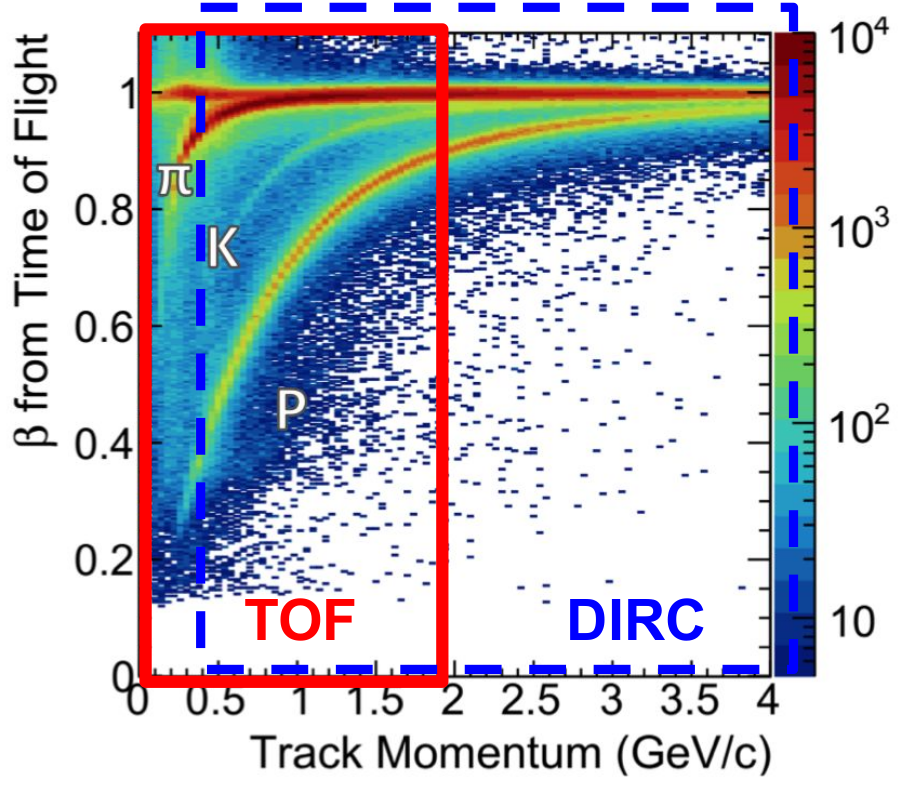}     
 	\includegraphics[width=0.49\textwidth]{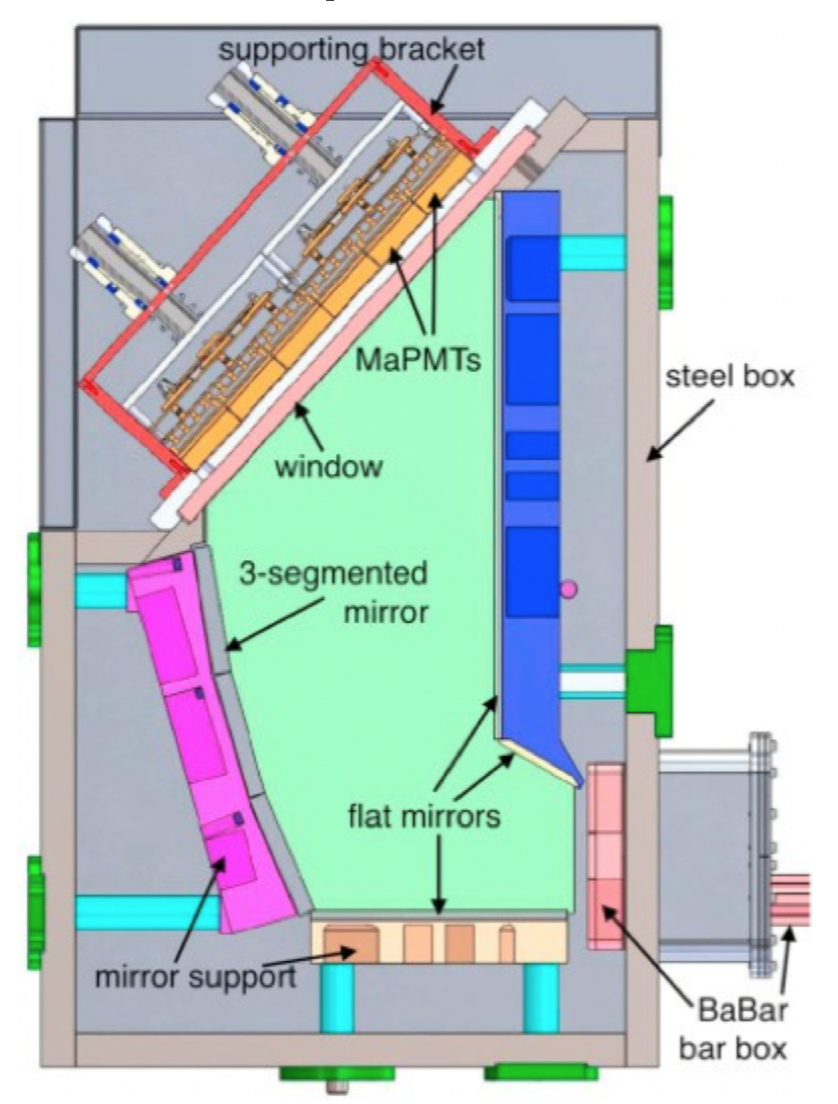}
    \caption{(a) Reconstructed $\beta$ vs track momentum distribution from $\gx{}$ Phase I data. The $\pi$, $K$ and proton trajectories are labelled. The TOF PID capability is indicated by the red box; the expected PID capability after the DIRC integration is indicated by the blue box. (b). Layout of the DIRC optical box shows the locations of optical components, including: PMTs, mirrors, and fused silica barbox attachment points. Figure (b) was first published in Ref.~\cite{ali20}}
	\label{fig:pid}
\end{figure}

The \gx{} experiment, shown schematically in Fig.~\ref{fig:GlueXcartoon} is ongoing at Jefferson Lab Hall D. It utilizes a tagged photon beam derived from the electron beam's coherent bremsstrahlung radiation interacting with a thin diamond wafer.  The primary experimental objective is to search for and to study hybrid mesons, which contain intrinsic gluonic degrees of freedom in the construction of their wave functions~\cite{PAC30,PAC40,PAC42}. Hybrid meson states are predicted by Lattice QCD calculations~\cite{Dudek:2013yja}, and are considered as a quantitative test of our understanding of the strong nuclear force in the non-perturbative regime.


Construction and installation of the beamline components and the \gx{} Phase I (low-intensity period) detector configuration (excluding the DIRC) were completed in 2014. The first physics-quality photon beam was delivered in the Spring of 2016~\cite{Ghoul:2015ifw} and the Phase I of the \gx{} physics program run from 2017-2018. The particle identification (PID) capabilities have been studied with this data and in the forward region and the time-of-flight (TOF) detector performance has reached its design specifications for providing $\pi/K$ separation up to $p\sim2$~GeV$/c$, as indicated by the red box in Fig.~\ref{fig:pid}(a).  The analysis and search for hybrid mesons which decays to non-strange quark final states are currently ongoing.

For Phase II of the \gx{} experiment, an upgrade (the DIRC) is needed to enhance the PID capabilities to study the strange quark flavor content of hybrid states, thus exploiting the full discovery potential. The DIRC upgrade for \gx{}, described in detail in Ref.~\cite{dirc_tdr}, utilizes one-third (four modules) of the fused silica radiators from the decommissioned BaBar DIRC (Detection of Internally Reflected Cherenkov light) detector~\cite{Adam:2004fq}, attached to two newly constructed compact expansion volumes (known as Optical Boxes) which are equipped with pixelated detector read-out, see Fig.~\ref{fig:pid}(a). The DIRC upgrade will expand the PID capability beyond 2 GeV$/c$, up to 4 GeV$/c$. In these proceedings, we briefly describe the installation of the \gx{} DIRC and the detector commissioning.

\section{Detector Construction and Installation}
\label{sec:install}

\begin{figure} [th!]
    \begin{center}
        \subfloat[Unloading the first DIRC bar box]{\includegraphics[width=0.48\textwidth]{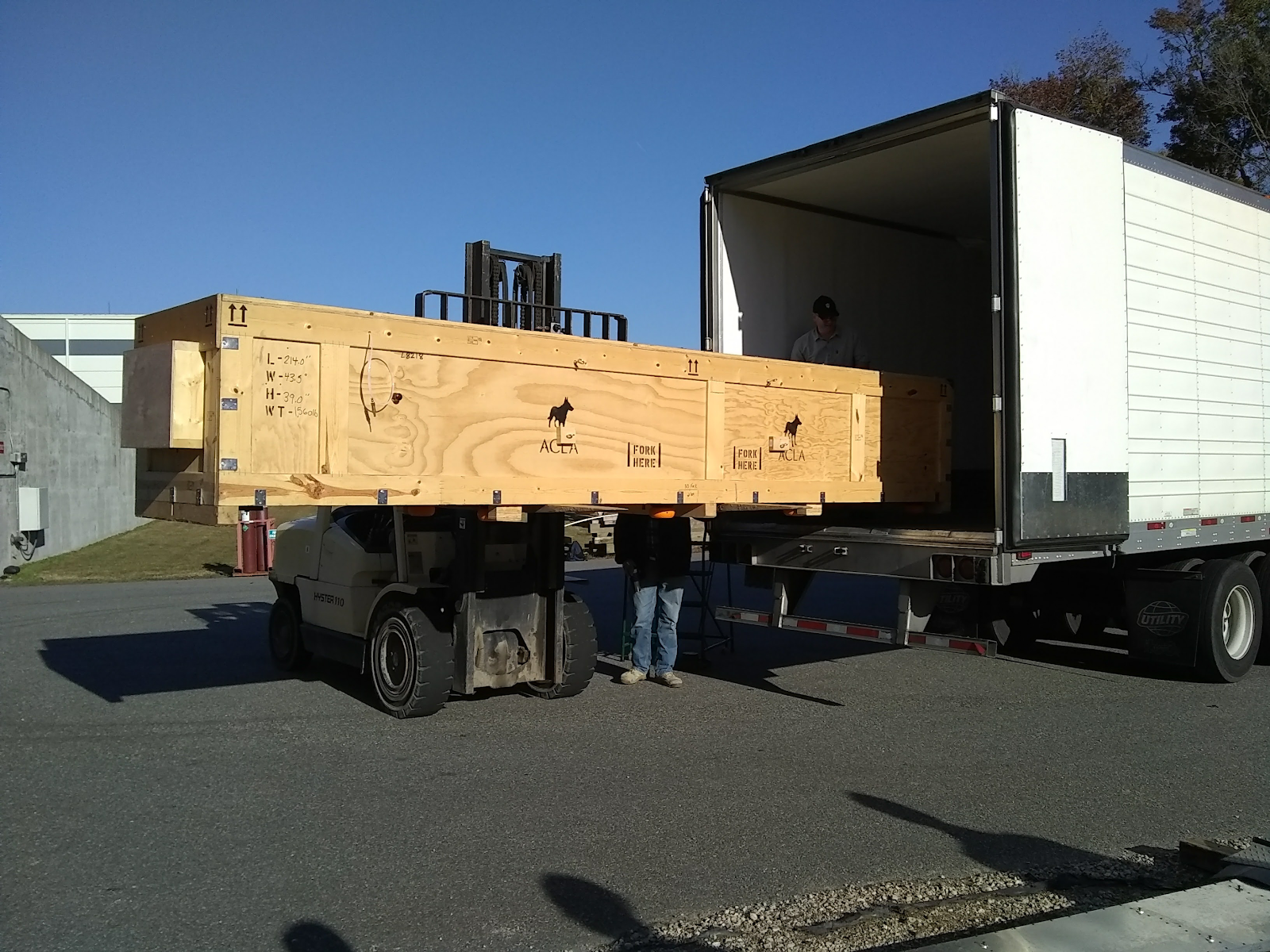}}~~
        \subfloat[All four bar boxes installed]{\includegraphics[width=0.48\textwidth]{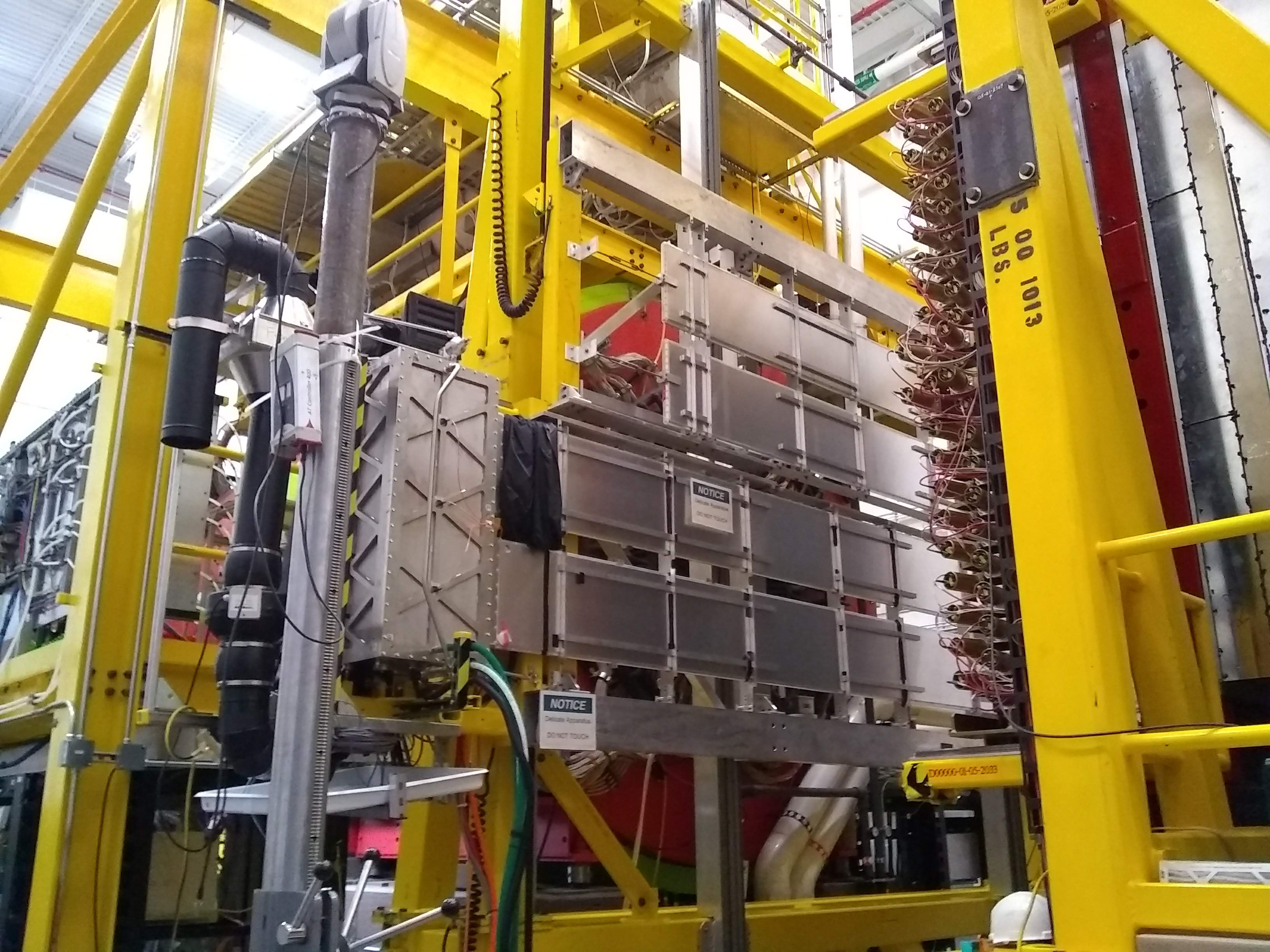}} \\ 
        \subfloat[Installed PMT modules]{\includegraphics[width=0.48\textwidth]{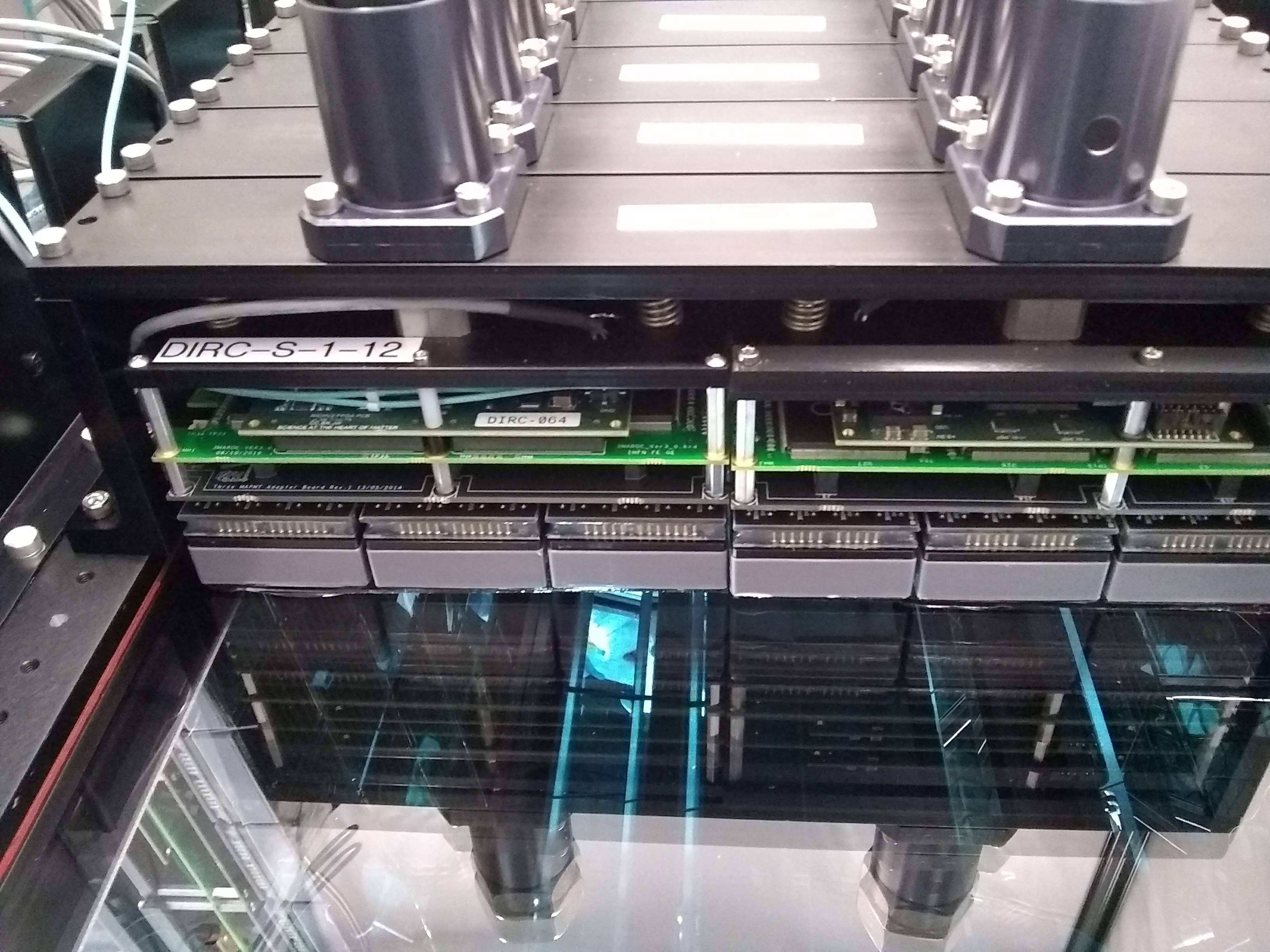}}~~
        \subfloat[PMT module stack layout]{\includegraphics[width=0.48\textwidth]{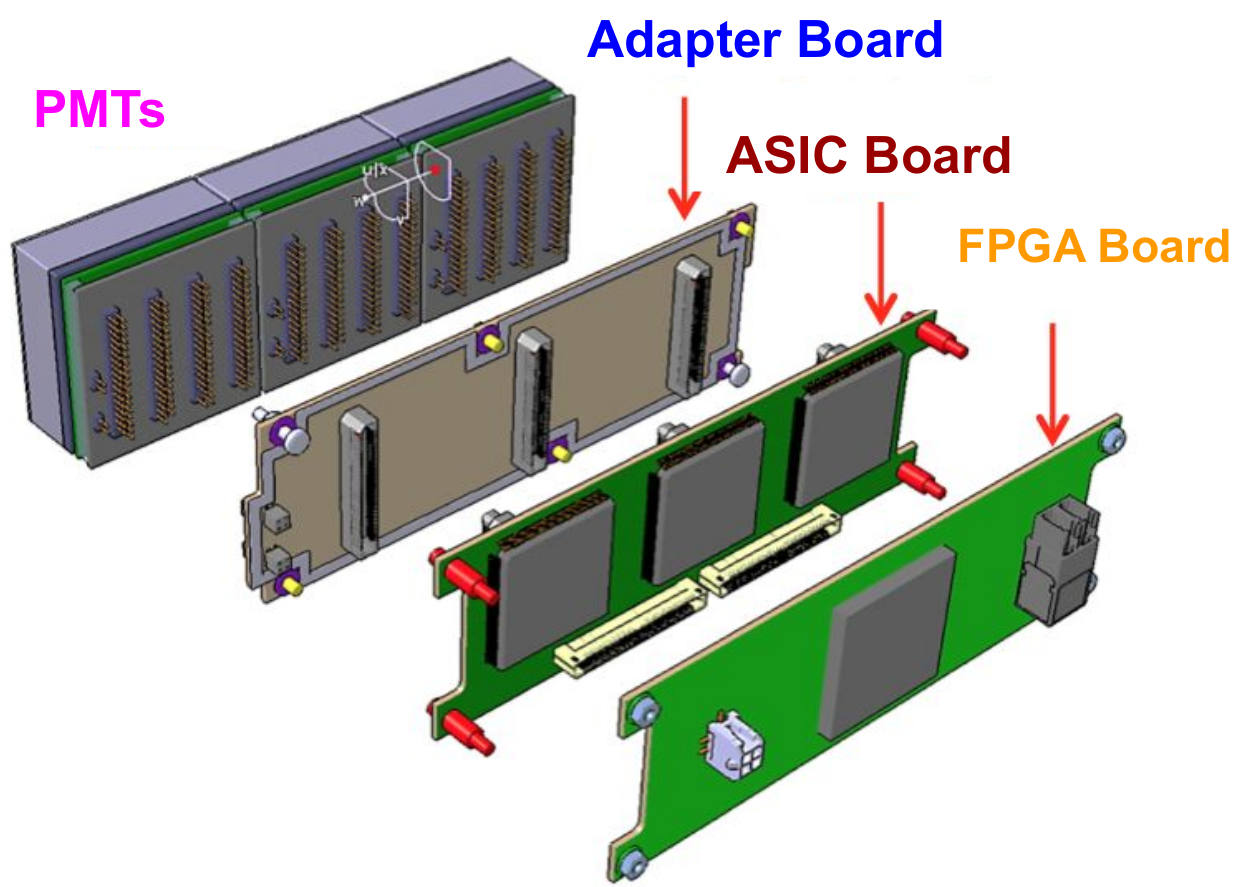}}
    \caption{(a) Unloading the first DIRC bar box at Jefferson Lab in Oct 2017. Note that the second shipment was for three DIRC bars. (b) Four BaBar bar boxes after transport to JLab and installed in the \gx{} detector on the DIRC support structure in Hall D. (c) and (d) show the layout of the DIRC PMT detector assembly modules and the attachment to the quartz window of OB. Figures (b) and (c) were first published in Ref.~\cite{ali20}.}
\label{fig:barbox}
\end{center}
\end{figure}

The first major task of the \gx{} DIRC installation project was the transportation of four DIRC ``bar boxes" from SLAC to JLab in 2018. These bar boxes were recycled components from the BaBar detector consisting of twelve 4.9~m long fused silica radiator bars, contained in an aluminum housing~\cite{Patsyuk:2020wnh}. The fragile nature of these bars made the transportation effort a major challenge. After careful evaluation, the bars were transported in two separate shipments by road with continuous careful monitoring. Customized transportation crates were designed and constructed.  Each hosted mechanical and air suspension system, accelerometers, an ultra pure N$_2$ purge system and cameras to monitor and ensure safety. During each transportation shipment, members of the DIRC transportation team constantly monitored parameters such as the changes in g-force, temperature, N$_2$ purging pressure and live images. For a detailed discussion of the transportation system see Ref.~\cite{hardin:2019}. Once at JLab, these bar boxes were installed in their support structure for operation in Hall~D, as shown in Fig.~\ref{fig:barbox}.

The expansion volume and photon collection component for the DIRC, known as the Optical Box (OB), consist of multiple flat aluminized mirrors, immersed in water, that reflect the Cherenkov photons through a fused silica window toward a plane of Multi-Anode PMTs (MAPMTs). The inner lay-out of the OB is shown in Fig.~\ref{fig:pid}(b). Note that water is used inside of the expansion volume to approximately match the refractive index of fused silica and avoid photon losses.

The front surface, pre-aluminized flat mirrors\footnote[1]{https://firstsurfacemirror.com} have an optical reflectivity of $>$ 90\% for the wavelength range of 200-400 nm. Note that the photon detection efficiency of the DIRC system drops significantly at wavelength below 300 nm. This is primarily due to the borosilicate face of PMTs, Epotek 301-2 glue\footnote[2]{http://www.epotek.com/site/administrator/components/com\_products/assets/files/Style\_Uploads/301-2RevXII.pdf} (for the fused silica bars) and water absorption (purity dependent). To ensure the photon transmission quality, the water in each of the OB ($\sim$450 liters) is constantly circulated at a rate of 10 times per day. 

The Hamamatsu H12700 MAPMTs~\footnote[3]{https://www.hamamatsu.com/jp/en/product/type/H12700B/index.html} are used as the photon detector. The effective area of a single MAPMT is 48.5 mm $\times$ 48.5 mm (square), and is segmented in 64 individual pixels. The evaluation of the performance of the MAPMTs and the integrated electronics readout system was performed at JLab. The MAPMT test used a dedicated laser test setup which illuminates the effective detection area to measure the single photoelectron response for each pixel individually to determine the gain and efficiency, following the procedure described in Ref.~\cite{Contalbrigo:2020snw}.

%



Each OB is equipped with 36 PMT module assemblies: mounting 90 PMTs and 18 plastic square dummy PMTs. The total number the DIRC readout channels for both OB are 11520. The MAPMTs are coupled to the silica window of the OB by a UV transparent~\cite{li14} silicone cookie (Momentive RTV 615 \footnote[4]{https://www.momentive.com/en-us}) and compressed by the anodized aluminum bracket, as shown in Fig.~\ref{fig:barbox}(c).

Each PMT module assembly consists of three PMTs, a high voltage adaptor board, an ASIC board, and a FPGA Board, as shown in Fig.~\ref{fig:barbox}(d). The application-specific integrated circuited (ASIC) board hosts three (one per each PMT) MAROC chips, which amplify and digitize the current pulse from the single photon hits on each MAPMT pixel, then deliver the hit time as well as the time-over-threshold information; the processed information (pixel-by-pixel) are then transferred to the trigger supervisor modules through the field-programmable gate array (FPGA) board over a fiber link. During the commissioning, we varied the HV settings, the MAROC gain settings and thresholds, to determine the optimum operating point. At the \gx{} Phase II production condition, DIRC system data transfer speed can reach up to 700 MegaBytes/s.

Each on-board FPGA operates at 5V with a current draw of 1 A, the heat power output inside the closed PMT dark box sums to 200 W. For heat dissipation purposes, a blower fan which is capable of moving enough air through a heavy duty filter was introduced to keep the FPGA operating temperature below 60$^\circ$C, which is the Artix 7 chipset\footnote[5]{https://www.xilinx.com/products/silicon-devices/fpga/artix-7.html} recommended temperature limit for long term operation.

During the nominal operation, the MAPMTs in each OB are directly illuminated by three blue LED sources at 405 nm in wavelength. In order to minimized the secondary scattered photon, each LED source is diffused into a square pattern by an engineering diffuser with an opening angle of 50 degrees (in air) to cover one third of the detection plane. The LEDs are pulsed at a frequency of 150 Hz and the three LED bunches are 10 ns apart. These valuable data are not only useful for the live calibration, they also offer important insight for the long term performance of the detector system, including water quality, detector efficiency, and data acquisition.

\section{Preliminary Commissioning Results}

In 2019, two separate DIRC commissioning periods took place: the first was the ten-day run in February 2019, when only half of the \gx{} DIRC was commissioned at the nominal \gx{} Phase I experimental condition; the second period lasted for two weeks in December 2019, where the full detector assembly was commissioned and the operation at \gx{} Phase II (high intensity) condition was also tested. Roughly 10 Billion triggered events were collected in each run period.



\begin{figure} [t]
    \centering
    \includegraphics[width=0.49\textwidth]{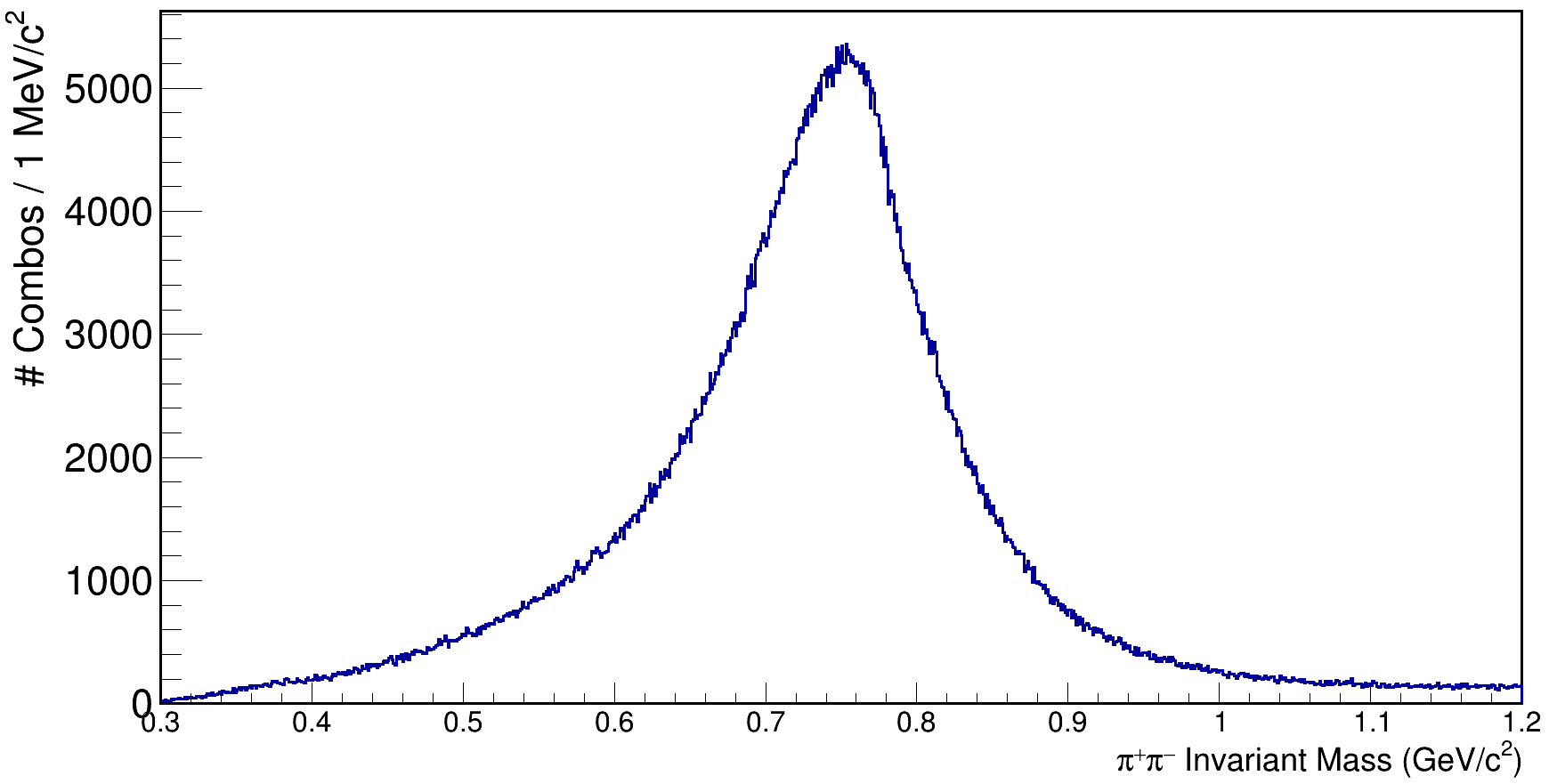}
    \includegraphics[width=0.49\textwidth]{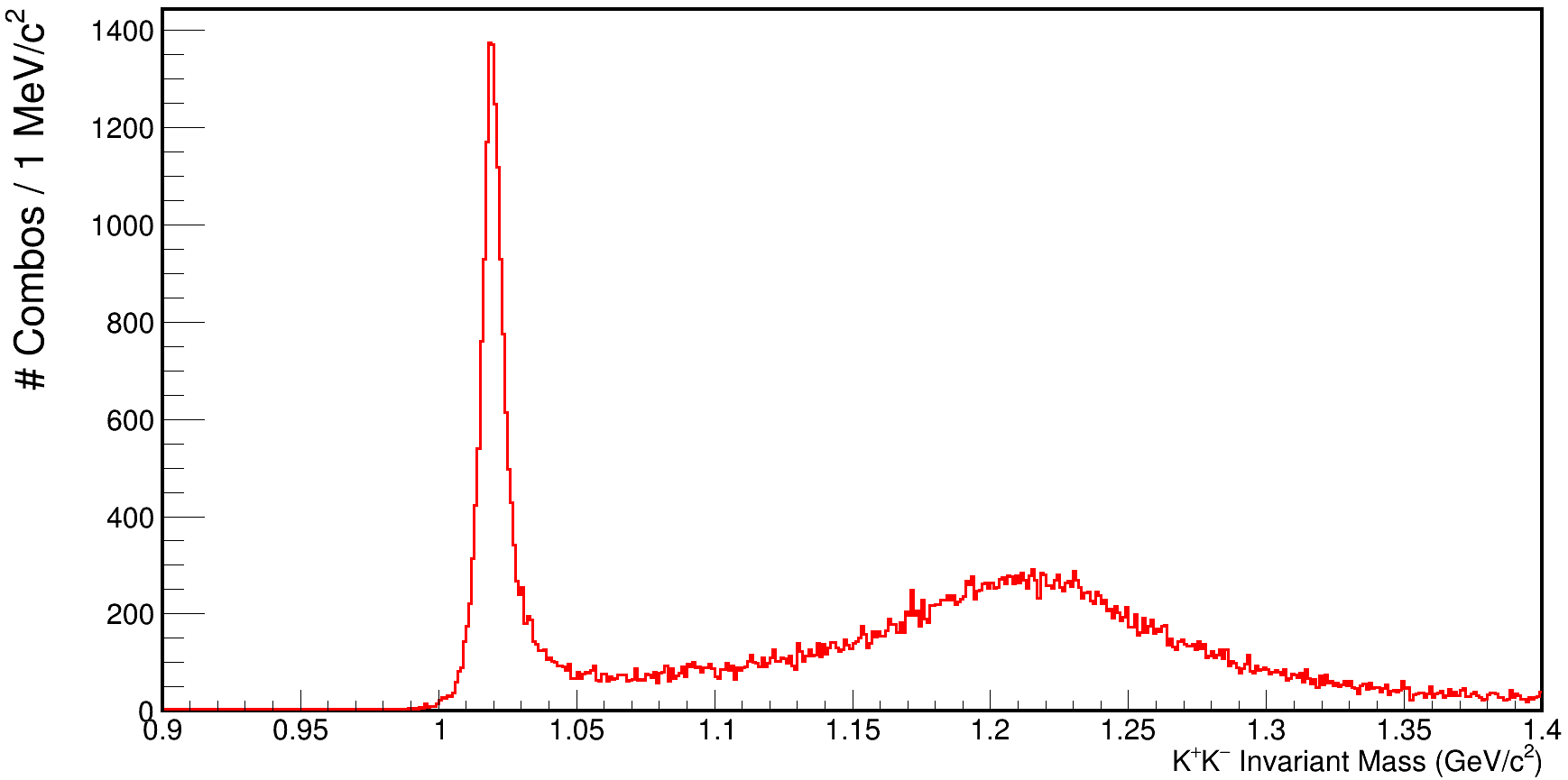}
    \caption{Invariant mass of exclusively produced $\pi^+\pi^-$ (left) and $K^+K^-$ (right) showing clear $\rho$ and $\phi$ peaks selected to study the DIRC detector performance. A 150 MeV window (675-825 MeV) is applied to select the $\rho$ events and a 50 MeV (990-1140 MeV) window is applied for $\phi$.
}
    \label{fig:rhophi}
\end{figure}

Under the nominal \gx{} Phase I experimental condition, the final state particle reconstruction yields large statistical samples of exclusive $\rho$ ($\gamma p \rightarrow \rho p,  \rho \rightarrow \pi^+\pi^-$) and $\phi$ ($\gamma p \rightarrow \phi p,  \phi \rightarrow K^+K^-$) events, as shown by the invariant mass distributions in Fig.~\ref{fig:rhophi} (left) and (right), respectively.  These events provided pure samples of $\pi^\pm$ and $K^\pm$ which were used to test the DIRC reconstruction algorithms and study the PID performance with the commissioning data.



\begin{figure} [t]
    \centering
    \includegraphics[width=0.70\textwidth]{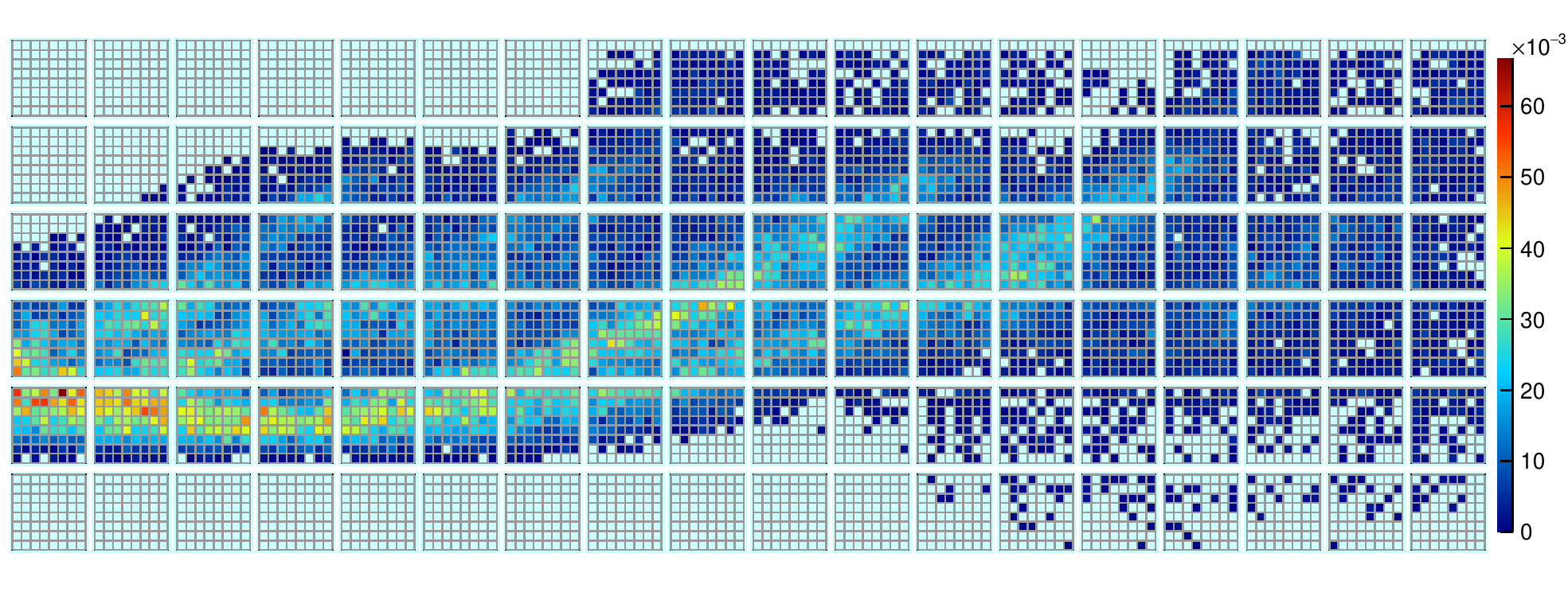}
    \includegraphics[width=0.70\textwidth]{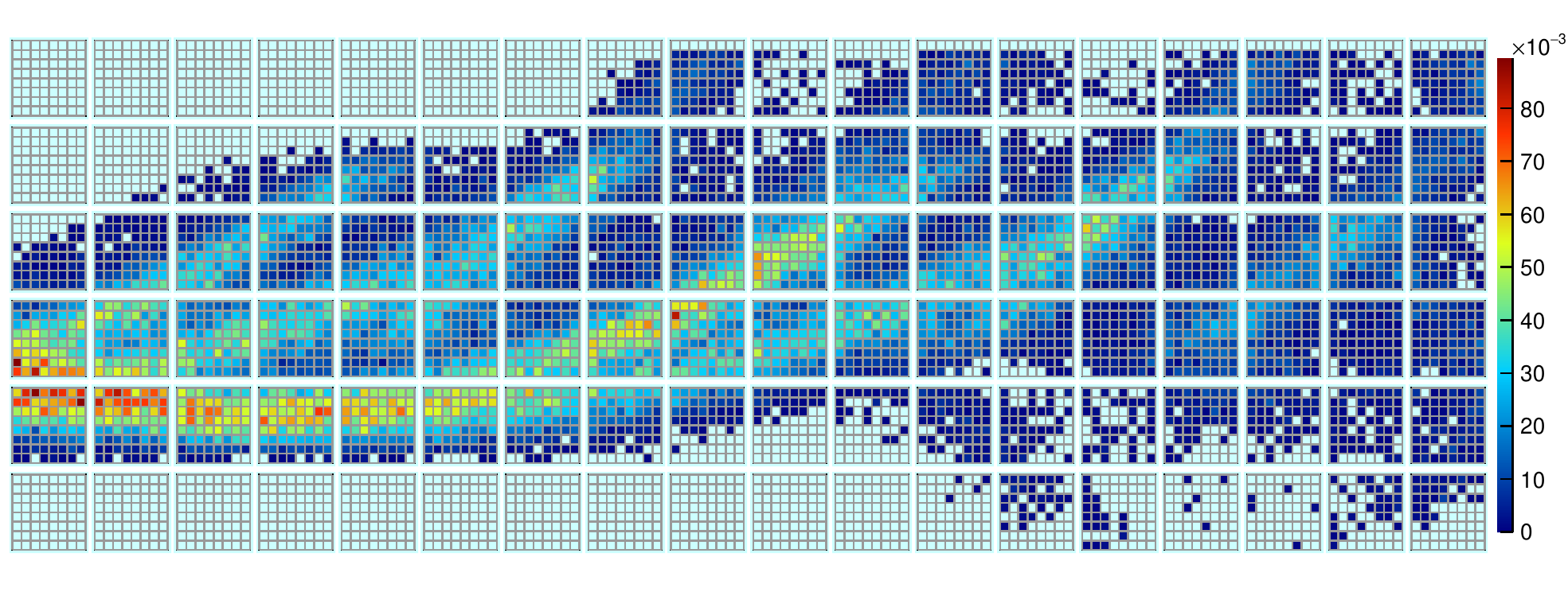}
    \caption{Cherenkov photon hit pattern on the MAPMT plane from the DIRC South (lower) OB for a sample of identified $\pi^+$ tracks, comparing data (top) with expected distribution from Geant 4 simulation (bottom). These images are rotated by 90 degrees counterclockwise. In low occupancy regions, there are seven dummy PMTs located in the top column (counting from the left) and eleven in bottom column. The hit pattern and the data-simulation agreement for the North (upper) OB is similar. Figures first published in Ref.~\cite{ali20}.}
    \label{fig:hit_pattern} 
\end{figure}

The photon detection plane of each OB consists of 18 rows $\times$ 6 columns of MAPMTs and dummy PMTs (144 rows $\times$ 48 columns of pixels) that detects the images of Cherenkov rings from the DIRC bar. Fig.~\ref{fig:hit_pattern}(a) shows the accumulated hit pattern histograms (rotated by 90 degree) that were generated by a small sample of $\pi^+$ tracks for the South (lower) OB; the tracks from the data are shown on the upper panel and simulated distribution (using Geant 4) are on the bottom panel. We used a detailed Geant4 simulation, which included the wavelength-dependent material properties, the photon transport efficiency (based on measurements) in the bars, the mirror reflectivity, and the photon detection efficiency. The excellent agreement of the occupancy distributions shows that the geometry of the bars, mirrors, and MAPMTs is modelled correctly in the simulation.




\begin{figure} [t]
    \centering

    \subfloat[Cherenkov angle]{\includegraphics[width=0.49\textwidth]{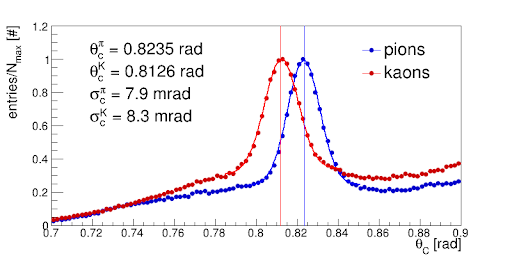}}
    \subfloat[Computed likelihood distributions]{\includegraphics[width=0.49\textwidth]{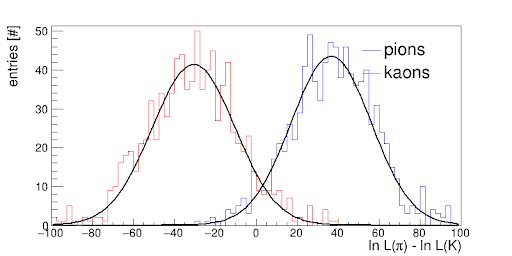}}
    \caption{(a) Cherenkov angle $\theta_C$ distribution per photon for identified pions (blue) and kaons (red) within a momentum range 2.7-3.3~GeV/$c$ identified through the $\rho$ and $\phi$ reactions. (b) Computed log-likelihood difference distributions for $\pi$ (in blue) and $K$ (in red) mass hypothesis. These results were from late 2019.}
    \label{fig:sep_power}

\end{figure}

The essence of DIRC PID algorithm involves comparing the observed Cherenkov photon pattern for a single particle track to the expected Cherenkov angles for a $\pi$ and $K$ mass hypothesis. Then the likelihood for each hypothesis is computed assuming a Gaussian resolution on the Cherenkov angle. The measured Cherenkov angles per photon for charged pions and kaons are from the $\rho$ and $\phi$ calibration samples are shown in Fig.~\ref{fig:sep_power}(a) for particles with a momentum between 2.7 and 3.3 GeV/c momentum.


A central location of the DIRC bar box directly above the photon beam path is selected as the study sample. Here, a good (data-simulation) agreement was observed in the Cherenkov angle $\theta_C$ and single photon resolution ($\sigma_C$). The difference between $\pi$/$K$ likelihood distributions is presented in Fig.~\ref{fig:sep_power}(b), the resulting separation power is $>$3$\sigma$, which meets the expected performance. The global $\pi/K$ separation power extraction is currently underway. The average number of detected photons per track ranges from 15 to 35, depending on the track incident angle. This is consistent with the expected photon yield derived from the BaBar operation experience.  




\section{Operation Experience}

\begin{figure} [th]
    \centering
    \includegraphics[width=1\textwidth]{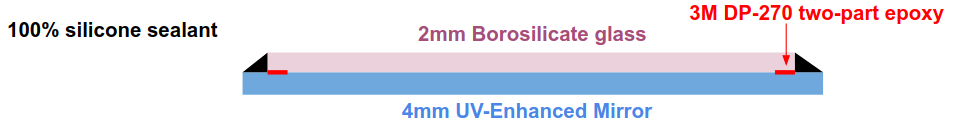}
    \caption{Section view of the mirror configuration. }
    \label{fig:mirror_config}
\end{figure}

During the first DIRC commissioning in spring 2019 the mirrors were exposed to water for more than four months, and the reflectivity of the aluminized mirror (even by a visual inspection standard) was found to be degraded. Two major issues were identified:
\begin{itemize}
\item A careful study and inspection revealed that the reflective coating surface developed small pitted holes when directly exposed to water, these holes became optically transparent. 
\item The reflective coating surface developed irremovable long white trail marks (haze), which caused up to a 80\% loss of mirror reflectivity, see Ref.~\cite{wenliang2019} for full detail. 
\end{itemize}

These findings motivated the DIRC construction team to develop a new strategy for operating mirrors under water. The photon losses have to be avoided to improve the PID performance of the DIRC PID algorithm, because every lost photon (compared to the expectation) will reduce the $\pi/K$ separation power. Therefore, protecting the mirror surfaces from corrosion during operation is critical. 

The new mirror configuration (Fig.~\ref{fig:mirror_config}) involved a sheet of borosilicate glass to act as a protector; clear epoxy (3M DP270~\footnote[2]{https://multimedia.3m.com/mws/media/66773O/3mtm-scotch-weldtm-epoxy-potting-compound-adhesive-dp270.pdf}) was used to bond the mirror and the glass protector along the edges (epoxy bonding width is $<$3 mm); while a bead of 100\% silicone was applied as sealant against water. The loss in reflective area of the smallest mirror (the worst case scenario) was 5\%, after the configuration was applied. The visual inspection after the 2019 commissioning confirmed the new mirror configuration was a success in protecting the mirror coating against the water corrosion.

Additionally, once per month during the DIRC operation period, the OB water sample (directly collected from the box) transmission quality was measured in the wavelength range of 300-600~nm and no evidence of degradation in photon transmission was observed.


\section{Conclusion and Outlook}

In these proceedings, we reported on the installation and commissioning success of the DIRC detector in February and December, 2019. The integration of the DIRC system marked the beginning of the $\gx{}$ experiment Phase II running. The analysis of the commission data is ongoing, however the initial results obtained for the small commissioning data set already show that the obtained hit patterns and Cherenkov angle resolution are consistent with the expectation and with Geant4 simulation.  We expect the measured performance to improve further once advanced calibration and alignment information is applied.



\section*{Acknowledgments}

We would like to acknowledge the outstanding efforts of the staff of the Accelerator and the Physics Divisions at Jefferson Lab that made the experiment possible.  This work is supported by the U.S. Department of Energy, Office of Science, Office of Nuclear Physics under contracts DE-AC05-06OR23177, DE-FG02-05ER41374 and Early Career Award contract DE-SC0018224 and the German Research Foundation, GSI Helmholtzzentrum f\"ur Schwerionenforschung GmbH.

\end{document}